\documentclass[aps,prb,manuscript]{revtex4}
\usepackage{graphicx}
\usepackage{epsfig}
\usepackage{color}
\usepackage{bm}
\usepackage{float}

\begin{document}
\draft
\title {Magneto-transport properties of doped graphene}

\author{Po-Hsin Shih$^{1}$, Thi-Nga Do$^{2,3}$, Godfrey Gumbs$^{4,5}$, Danhong Huang$^{6}$, and Ming-Fa Lin$^{1}$}
\affiliation{$^{1}$Department of Physics, National Cheng Kung University, Tainan, Taiwan 701\\
$^{2}$ Laboratory of Magnetism and Magnetic Materials, Advanced Institute of Materials Science,
Ton Duc Thang University, Ho Chi Minh City, Vietnam\\
$^{3}$ Faculty of Applied Sciences, Ton Duc Thang University, Ho Chi Minh City, Vietnam\\
$^{4}$Department of Physics and Astronomy, Hunter College of the City University of New York, 695 Park Avenue, New York, New York 10065, USA\\
$^{5}$Donostia International Physics Center (DIPC), P de Manuel Lardizabal,
 4, 20018 San Sebastian, Basque Country, Spain\\
$^{6}$US Air Force Research Laboratory, Space Vehicles Directorate, Kirtland Air Force Base, New Mexico 87117, USA
}

\date{\today}

\begin{abstract}

The effect due to doping by B, Si, N on the magneto-transport properties of graphene is investigated using the generalized tight-binding model in conjunction with the Kubo formula. The crucial electronic and transport properties are greatly diversified by the type of dopant and doping concentration. The contribution from the guest atoms may open a band gap, thereby giving rise to the rich Landau level energy spectra and consequently the unique quantum Hall conductivity. The Fermi energy-dependent quantum Hall effect appears as a step structure having both integer and half-integer plateaus. Doping leads to the occurrence of zero conductivity, unlike the plateau sequence for pristine graphene. The predicted dopant- and concentration-enriched quantum Hall effect for doped graphene should provide useful information for magneto-transport measurements and possible technological applications as well as metrology.

\end{abstract}
\pacs{PACS:}
\maketitle

\section{Introduction}
\label{sec1}

Understanding the physics behind the quantum Hall effect (QHE) has led to a prolonged effort by numerous theoretical and experimental researchers. In particular, the Hall conductivity of condensed matter systems under an external magnetic field has received special attention.\cite{1,2,3,4,5,6,7,8,9,10,11,12,13,14,15,16,17,18,19,20} Individual behaviors of magnetic-field-quantized Landau levels (LLs) have been employed to explain the QHE. The electrons filling LLs and wave functions have been demonstrated to be associated with the integer and fractional quantum Hall states. In this work, we will systematically explore the QHE in Si-, B-, and N-doped graphene for various doping concentrations with the aid of the tight-binding model to determine the energy bands.

\medskip
\par
Ever since the time when graphene was experimentally discovered, its exotic QHE has attracted many studies from both the experimental and theoretical points of view.\cite{1,2,3,4,5,6,7,8,9,10,11,12,13,14} Monolayer graphene shows amazing physical properties because of its single-atom thickness and hexagonal symmetry. The magnetically quantized LLs of the isotropic Dirac cone form a distinctive energy spectrum for which conduction and valence energy levels are dependent on both the field strength and the quantum number of specified LLs. Such an established relation, $\sqrt{B_zn^{c,v}}$, has been verified through experiment using scanning tunneling spectroscopy (STS),\cite{21} optical spectroscopies,\cite{22} and transport equipment.\cite{5,6} Interestingly, it has been realized that the inter-LL transitions obey a specific selection rule associated with the equivalence of the two sublattices in the crystal structure. Only excitations between the valence and conduction LLs whose quantum numbers satisfy ${\Delta\,n=|n^v-n^c|=1}$ are allowed, resulting in the exotic QHE. Magnetic transport measurements have verified the unconventional half-integer Hall conductivity $\sigma_{xy}=(m+1/2)4e^2/h$ ($m$ is an integer and the factor of $4$ represents the spin and sublattice-dependent degeneracy) in monolayer graphene.\cite{5,6} Such an extraordinary quantization phenomenon is related to the quantum anomaly of the ${n=0}$ LL coming from the Dirac point.\cite{1}

\medskip
\par
Currently, scientists are in search of novel materials yielding unusual physical phenomena and having potential for applications. By generating graphene-based defect systems, such as substituting impurities or guest atoms in a hexagonal carbon lattice, ones could effectively manipulate the fundamental properties inherent to graphene. Among various pertinent guest atoms, Si,\cite{23} B,\cite{24, B} and N\cite{25, N} have been successfully substituted for the carbon host atoms by chemical vapor deposition (CVD) or arc discharge experiments. The doping effect eliminates the equivalence of the two sublattices, causing dramatic changes in physical properties, including the energy gap opening and the deviation from the  original Dirac cone. Previous studies have proved the emergence of different ionization potentials and non-uniform hopping integrals.\cite{26,27} Specifically for Si-, B-, and N-doped graphene, the $\pi$ bonding extending on a hexagonal lattice is significantly distorted. As a consequence, the low-lying energy band is controlled by both the dopants and C atoms simultaneously. The magnetically quantized LLs of the dopant- and concentration-created unusual band structures have been theoretically predicted to be feature-rich and unique.\cite{28} These facts make it promising for the QHE of such special LL energy spectra to yield abundantly rich results.

\medskip
\par

The rest of our paper is outlined as follows. In Sec.\ \ref{sec2}, we employ the tight-binding method to solve for the eigenvalues and eigenfunctions of the magnetic Hamiltonian, and then use these results to calculate the Hall conductivity by means of the linear Kubo formula. This procedure enables us to identify allowed transition channels through the aforementioned magneto-electronic selection rules, leading to the understanding of Fermi-energy- and magnetic-field-diversified quantum conductivity. Fundamental properties of doped graphene are greatly enriched by various types of dopants and their concentrations. In Sec.\ \ref{sec3}, we investigate the QHE of several dopants of Si, B, and N with 2\% and 12.5\% of doped atoms. Our results demonstrate the robust correlation between the unusual QHE and the dopant- and concentration-induced rich LLs in the doped graphene systems. Our theoretical predictions provide essential information for future experimental verification of the QHE in graphene-based defect materials. Section \ref{sec4} is devoted to our concluding remarks.

\section{Method}
\label{sec2}

The crystal structures of doped graphene for various concentrations are presented in Figs. 1(a) and 1(b). For a 1:2$n^2$  concentration doping ($n$ is the cell multiplicity), the super cell could be obtained by expanding the original unit cell of pristine graphene $n$ times along the lattice vectors $\vec{a_1}$ and $\vec{a_2}$, that is, ($n\vec{a_1}, n\vec{a_2}$). In this work, we choose a rectangular super cell as marked by the red lines for convenience in considering the effect of magnetic field. A single supercell comprises 2$n^2$ sublattices of two types, ${A_i}$ and ${B_i}$, in which $i$ indicates the lattice site. Both C-${2p_z}$ orbital and the guest orbitals of ${B-2p_z}$, ${Si-(2p_z, 3p_z)}$, and ${N-2p_z}$ play the key roles in governing the crucial characteristics at low-energy range. The Hamiltonian matrix includes the non-uniform bond lengths, site energies and nearest-neighbor hopping integrals, which could be written as follows:\cite{28}

\begin{eqnarray}
H_{j+2nl-2n,j+2nl-n}=H_{j+2nl-n,j+2nl-2n}^*=t_{j+2nl-2n,j+2nl-n}f_1  \nonumber  \\
H_{j+2nl-2n,m(j)+2nl-n+1}=H_{m(j)+2nl-n+1,j+2nl-2n}^*=t_{j+2nl-2n,m(j)+2nl-n+1}f_2 \nonumber \\
H_{j+2nl-2n,j+2n[m(l)+1]-n}=H_{j+2n[m(l)+1]-n,j+2nl-2n}^*=t_{j+2nl-2n,j+2n[m(l)+1]-n}f_3  \nonumber  \\
H_{1,1}=\epsilon_{B/Si/N-C},
\end{eqnarray}
in which, the hopping term is defined as

\begin{eqnarray}
t_{\alpha, \beta}=\left\{\begin{array}{ll}
                 \gamma_{B/Si/N-C}, \mathrm{if}\hspace{0.5em} \alpha\hspace{0.5em} \mathrm{or}\hspace{0.5em}  \beta\hspace{0.5em} \mathrm{equal}\hspace{0.5em}1\\  %
                 \gamma_{C-C}, \mathrm{otherwise}.
                \end{array} \right.
\end{eqnarray}
In this notation, $f_{1,2,3}$ is defined in terms of the wave vector $\vec{k}$ and the vectors connecting the nearest-neighbor lattice sites $\vec{R}_{1,2,3}$ through $e^{\mathrm{i}\vec{k}.\vec{R}_{1,2,3}}$. $m(k)=k+n-2$ mod $n$ is modulo function, and $j,l$ are the integers ($j,l=1,2,3,...,n$).

\begin{figure}[!h]
\centering
{\includegraphics[width=0.7\linewidth]{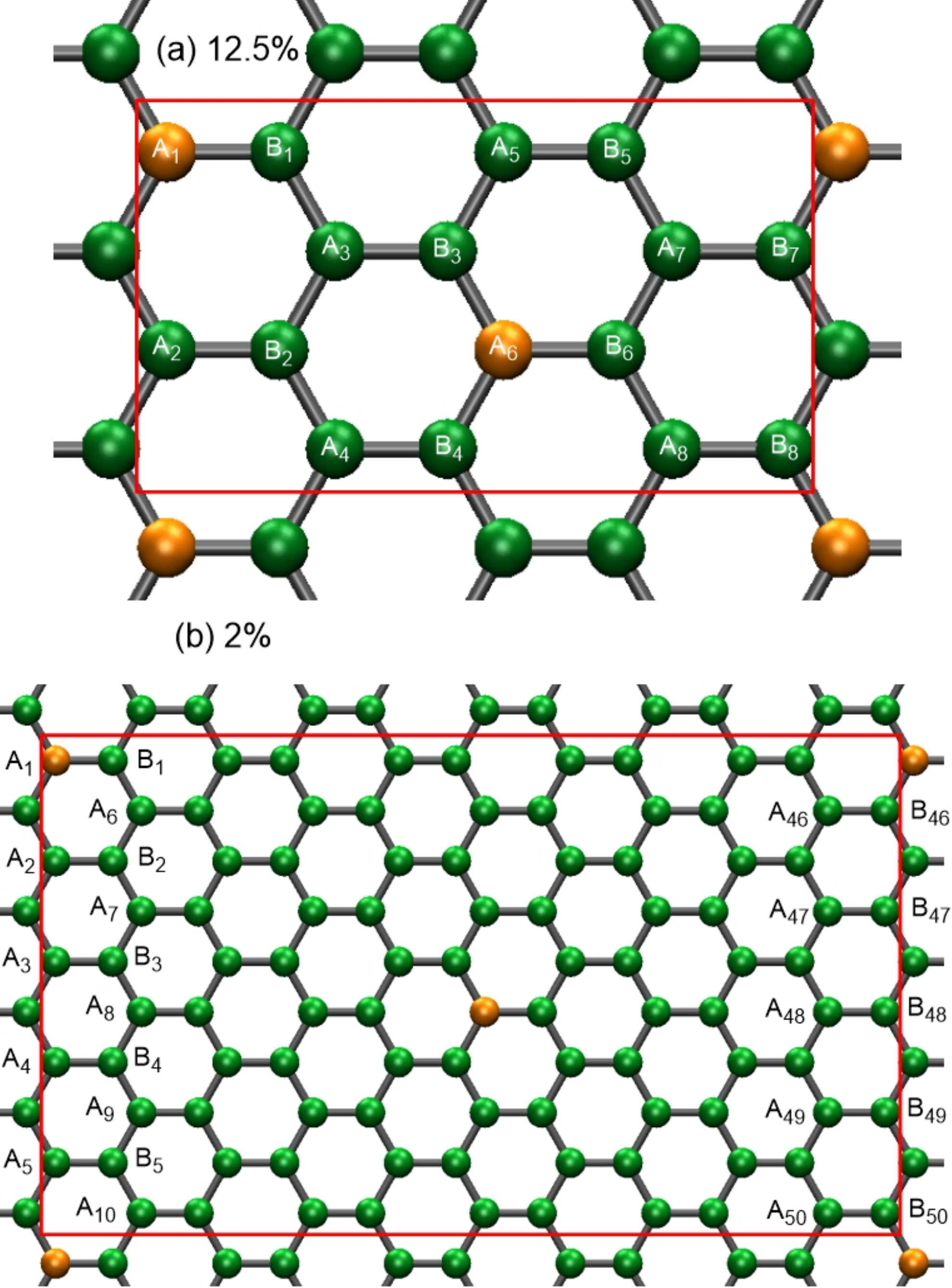}}
\caption{(Color online) Crystal structures of doped graphene for chosen doping concentrations of (a) 12.5\% and (b) 2\% dopants. The green and orange spheres denote, respectively, the C and Si/B/N atoms.}
\label{Figure 1}
\end{figure}

\medskip
\par

The C-C and B/Si/N-C bond lengths have been verified as being slightly different due to doping, especially the buckling effect of the guest atoms in graphene.\cite{26,27} For Si dopant, the guest atoms are located at a distance of $d_{Si}$ above graphene sheet, implying the significant adjustment of the $\pi$ bonding on a hexagonal lattice. The consequential non-uniform nearest-neighbor hopping integrals and site energies associated with the major ${2p_z}$ orbitals of the C host atoms and the minor ${2p_z/3p_z}$ orbitals of the guest atoms lead to extraordinary electronic properties. The parameters are appropriately fitted, as shown in Table I, so that the energy bands calculated by the tight-binding and first-principles methods are consistent.

\medskip
\par

\begin{table}[h]
\caption{Tight-binding parameters for pristine and doped graphene} %title of the table
\centering % centering table
\begin{tabular}{|c|c|c|c|c|} % creating 5 columns
\hline %inserting double-line
Parameter &Pristine graphene &Si-doped &B-doped &N-doped \\ [2ex]
\hline % inserts single-line
bond lengths (\AA)& $b$ = 1.42 & $b_{C-Si} = 1.7$ & $b_{C-B}$ = 1.42& $b_{C-N}$ = 1.42\\ % Entering row contents
\hline
buckled distance (\AA)& $d$ = 0 & $d_{Si}$ = 0.93& $d_{B}$ = 0& $d_{N}$ = 0\\
\hline
hopping (eV)& $t$ =-2.7& $t_{C-Si} = -1.3$ & $t_{C-B}$ = -2.17& $t_{C-N}$ = -2.1\\ % [1ex] adds vertical space
\hline
site energy (eV)& $\epsilon$ = 0 & $\epsilon_{Si}$ =1.3& $\epsilon_{B}$ =  2.33& $\epsilon_{N}$ = -2\\
\hline % inserts single-line
\end{tabular}
\label{table1}
\end{table}

\medskip
\par

The effect of a uniform perpendicular magnetic field on the system could be treated with the use of the vector potential-created Peierls phase of ${\Delta{\bf G}_{mm'}=\frac{2\pi}{\phi_0}\int_{R_{m'}}^{R_m}{\rm\bf A}({\rm\bf r})\cdot {\rm d} {\rm\bf r}}$ (${\phi_0\,=hc/e}$ the flux quantum) in the nearest-neighbor hopping integral.\cite{28} Such a phase is position-dependent so that the unit cell is extended to become along the armchair direction.  Consequently, the magnetic Hamiltonian is a large matrix whose periods strongly depend on the density of guest atoms and vector potential. This  Hamiltonian matrix is efficiently solved using the diagonalization method to obtain the LL energy spectrum and sub-envelope functions.\cite{29} Interestingly, the amplitudes of sub-envelope functions on the distinct sublattices in an enlarged unit cell could be regarded as their spatial distributions, and they are used, therefore, to analyze the main features of the LL wave functions.

\medskip
\par

Within linear response theory, the Hall conductivity is computed from the Kubo formula.\cite{30}

\begin{eqnarray}
\sigma_{xy} = \frac {ie^2 \hbar} {S}
&\sum_{\alpha} \sum_{\alpha \neq \beta} (f_{\alpha} - f_{\beta})
\frac {\langle \alpha  |\mathbf{\dot{u}}_{x}| \beta\rangle  \langle \beta |\mathbf{\dot{u}}_{y}|\alpha \rangle} {(E_{\alpha}-E_{\beta})^2 + \Gamma_0^2} \ .
\end{eqnarray}
In this notation, $|\alpha\rangle$ and $E_{\alpha}$ are, respectively, the LL state with energy, $f_{\alpha,\beta}$ the Fermi-Dirac distribution functions, $S$ is the area of the supercell, and $\Gamma_0$ a broadening parameter. Furthermore, the velocity operators $\mathbf{\dot{u}}_{x, y}$ directly determine the allowable inter-LL transitions; they are evaluated from the gradient approximation .\cite{31} The defect effect in doped graphene might induce the unconventional quantum conductivities.

\section{Results and Discussion}
\label{sec3}

\begin{figure}[!h]
\centering
{\includegraphics[width=0.8\linewidth]{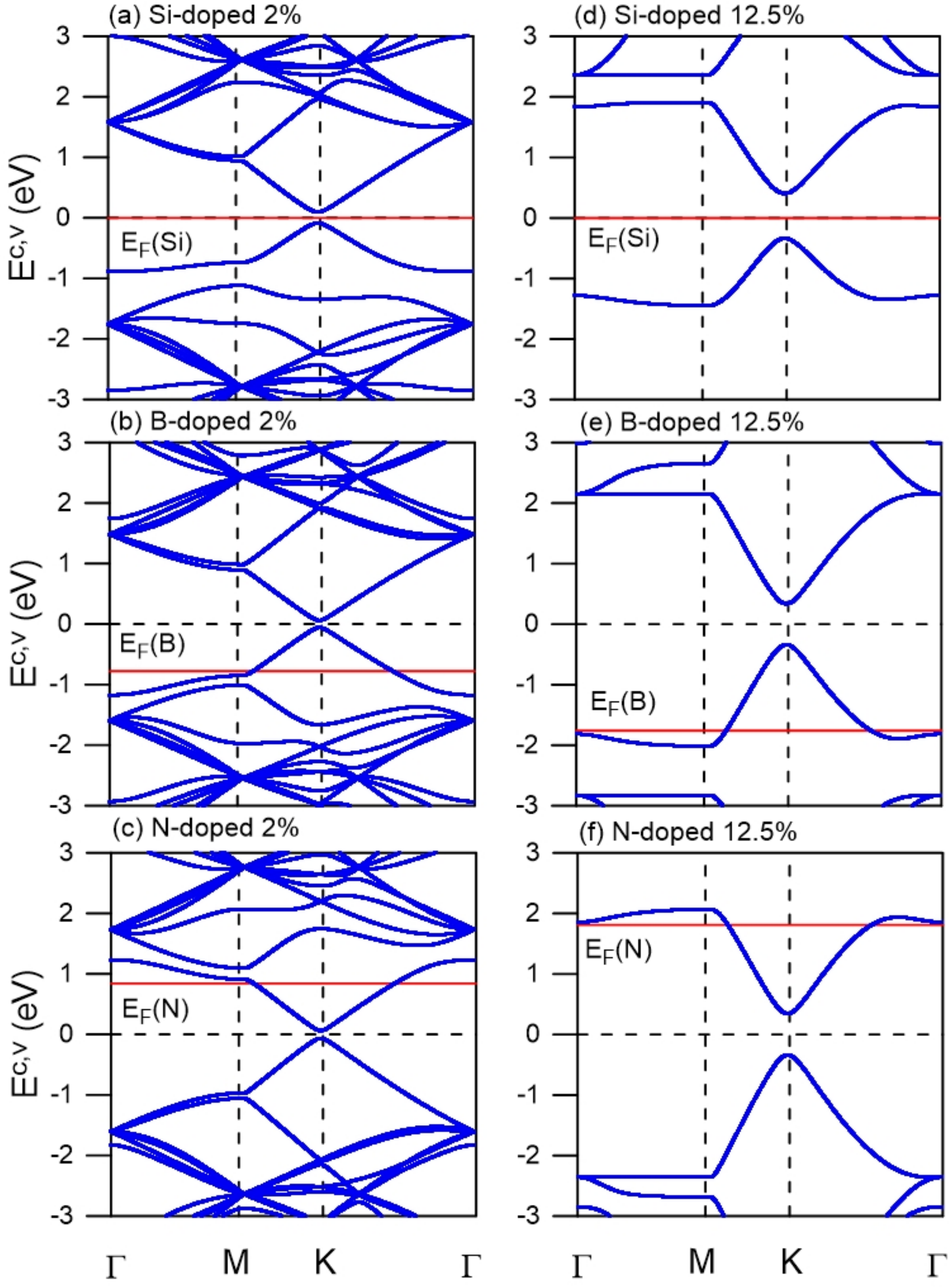}}
\caption{(Color online) Energy bands of doped graphene for Si, B, and N guest atoms with (a)-(c) 2\% and (d)-(f) 12.5\% concentrations, respectively.}
\label{Figure 2}
\end{figure}

The effect of doping with chosen dopants and concentrations remarkably diversifies the electronic properties of graphene. The substitution of guest atoms opens a direct band gap in monolayer graphene since it breaks the lattice symmetry, as illustrated in Figs. 2(a) through 2(f). The magnitude of the energy gap ($E_g$) is strongly determined by the dopants and doping concentration, as demonstrated in Table II for pristine graphene and Si-, B-, and N-doped systems with 2\% and 12.5\% guest atoms. Additionally, the significant influence of doping on the Fermi energy is worthy of consideration (Table III). For pristine graphene, the Fermi energy is equal to zero (Fig. S1 of the Supplemental Materials) and this remains the same in the presence of Si dopant. On the contrary, the Fermi energy levels are downshifted to the valence band or upshifted to the conduction band for B- and N-doped systems, respectively. These phenomena could be explained by the fact that a B atom acts as an p-dopant while an N atom is an n-dopant since a B/N atom has one less/more electron than a C atom. The 3D plots for energy bands with separation of occupied and unoccupied states are shown in Fig. S2 of the Supplemental Materials. Our numerical results are in agreement with previous theoretical calculations by the first-principles method.\cite{theo1,theo2,theo3} The prediction of the dopant- and concentration-dependent band structures in doped graphene should be useful for technological applications.

\begin{table}[h]
\caption{Band gap of doped graphene} %title of the table
\centering % centering table
\begin{tabular}{|c|c|c|c|} % creating 4 columns
\hline
$E_g$ (eV) &Si-doped &B-doped &N-doped \\ [2ex]
\hline % inserts single-line
2\% &  0.19 & 0.12 & 0.13\\ % Entering row contents
\hline
12.5\% & 0.74& 0.68& 0.68\\
\hline
\end{tabular}
\label{table2}
\end{table}

\begin{table}[h]
\caption{Fermi levels of doped graphene} %title of the table
\centering % centering table
\begin{tabular}{|c|c|c|c|} % creating 4 columns
\hline
Fermi levels (eV) &Si-doped ($E_F(Si)$) &B-doped ($E_F(B)$)&N-doped ($E_F(N)$) \\ [2ex]
\hline% inserts single-line
2\% &  0 & -0.78 & 0.84\\ % Entering row contents
\hline
12.5\% & 0& -1.76&  1.81\\
\hline % inserts single-line
\end{tabular}
\label{table3}
\end{table}

\begin{figure}[!h]
\centering
{\includegraphics[width=0.8\linewidth]{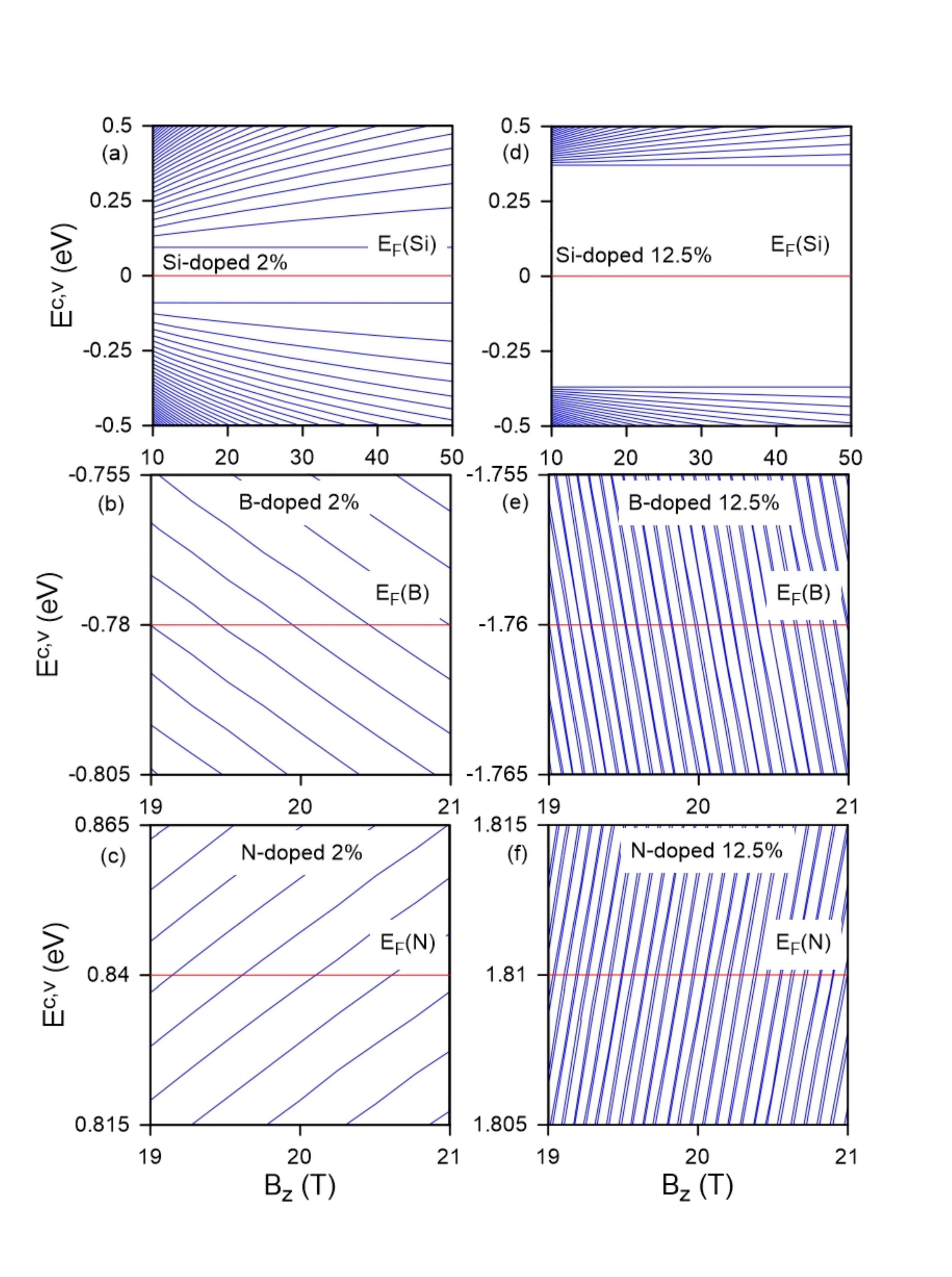}}
\caption{(Color online) The $B_z$-dependent LL energies of doped graphene with Si, B, and N dopants for (a)-(c) 2\% and (d)-(f) 12.5\%. }
\label{Figure 3}
\end{figure}

\medskip
\par

In the presence of a uniform perpendicular magnetic field, the quantized LLs show unusual characteristics due to the doping effect. Because of the opening of a band gap, the conduction and valence LL energy spectra are clearly separated, as shown in Figs. 3(a) and 3(d) for the magnetic-field-dependent energy spectra of Si-doped graphene.  The separation is invariant with  variation of the field strength so that it has no influence on the inter-LL transition between the valence and conduction bands. Each LL is four-fold degenerate, except for the lowest conduction and highest valence states in the vicinity of zero energy whose degeneracy is half smaller. For Si-doped graphene, the Fermi level is located at zero energy. Therefore, the former and the latter are unoccupied and occupied, respectively. On the other hand, they are either unoccupied for B or occupied for N dopants due to the shift of Fermi level. These two doubly degenerate LLs are split from the original zero-energy LL (n = 0) of pristine graphene due to the substitution of guest atoms. The $B_z$-dependence of LLs near the Fermi levels in B- and N-doped systems are worthy of detailed consideration. There, the density of LL states is very high, as clearly illustrated in Figs. 3(b) through 3(c) for 2\% and 3(e) through 3(f) for 12.5\% concentrations of B and N guest atoms. Especially for sufficiently high dopant density of 12.5\%,  the LLs show slight splitting behavior at higher and deeper energy ranges, altering the state degeneracy degree of freedom. These phenomena play a key role in the unique properties of quantum Hall conductivity, which we discuss below.

\begin{figure}[!h]
\centering
{\includegraphics[width=0.8\linewidth]{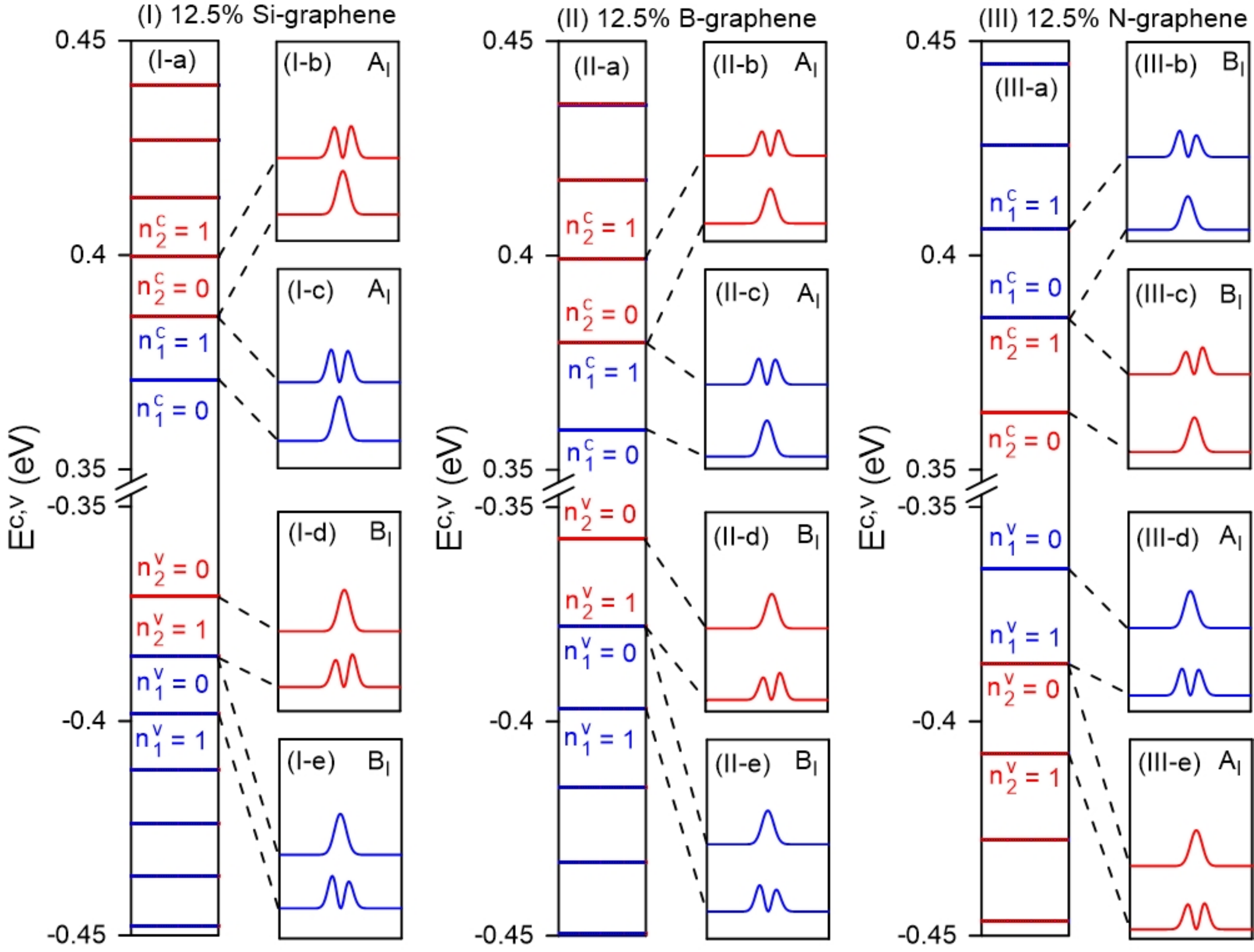}}
\caption{(Color online) The LL energies and wavefunctions of the first few levels for (I) Si, (II) B, and (III) N guest atoms with 12.5\% concentrations for $B_z$ = 20 T. The blue and red bars represent LLs localized at 1/6 and 2/6 centers, respectively.}
\label{Figure 4}
\end{figure}

\medskip
\par

The contribution of guest and host atoms to the Hall conductivity through the velocity matrix elements in Eq. (3) could be understood based on the amplitude of LL wave functions on the sublattices, as demonstrated in Figs. 4(I) through 4(III). Our numerical calculations show that the A sublattice which consists of the guest atoms obviously dominates the conduction bands of (Si, B)-doped systems and the valence band of N-doped graphene, and vice versa for the B sublattice. Especially for Si dopant with the presence of $3p_z$ orbital, its spatial distribution is even stronger than that of the C atoms. The quantum number, which is critical in understanding the selection rule of inter-LL transition, is determined based on the LL spatial distribution. Landau wave functions are only localized at some specific areas within the field-enlarged unit cell, particularly at 1/6, 2/6, 4/6 and 5/6 centers for a ($k_x$ = 0, $k_y$ = 0) state. LLs localized at 1/6 and 4/6 (2/6 and 5/6) are equivalent in the main characteristics and they are classified as $n^{c,v}_1$ ($n^{c,v}_2$) group. This is true for all the doped systems, regardless of the dopants and doping concentration. The doubly-degenerate conduction LL belongs to $n^{c}_1$ group for Si and B dopants but $n^{c}_2$ for N, while the opposite is true for the valence one.

\medskip
\par

The quantum Hall effect of doped graphene presents unusual and unconventional properties that may be manipulated by the degree to which the sample is doped. The Fermi energy-dependent conductivity quantization in an external magnetic field presents step structure, as illustrated in Fig. 5 for the resistivity $R_{xy} = 1/\sigma_{xy}$ of three different guest atoms with 2\% and 12.5\% concentrations. The plots of $\sigma_{xy}$ could be found in Fig. S3 of the Supplemental Materials. During the variation of the Fermi energy ($E_F$), a plateau is formed whenever one LL becomes occupied. The inter-LL transition between the occupied and unoccupied levels needs to satisfy a specific selection rule where the oscillation mode of the wave functions on the sublattices must be equivalent for the initial and final states. We observe that such a rule is $\Delta$n = $\pm$ 1 \& 0 for 2\% and $\Delta$n = $\pm$ 1 for 12.5\% doped graphene.

\begin{figure}[!h]
\centering
{\includegraphics[width=0.8\linewidth]{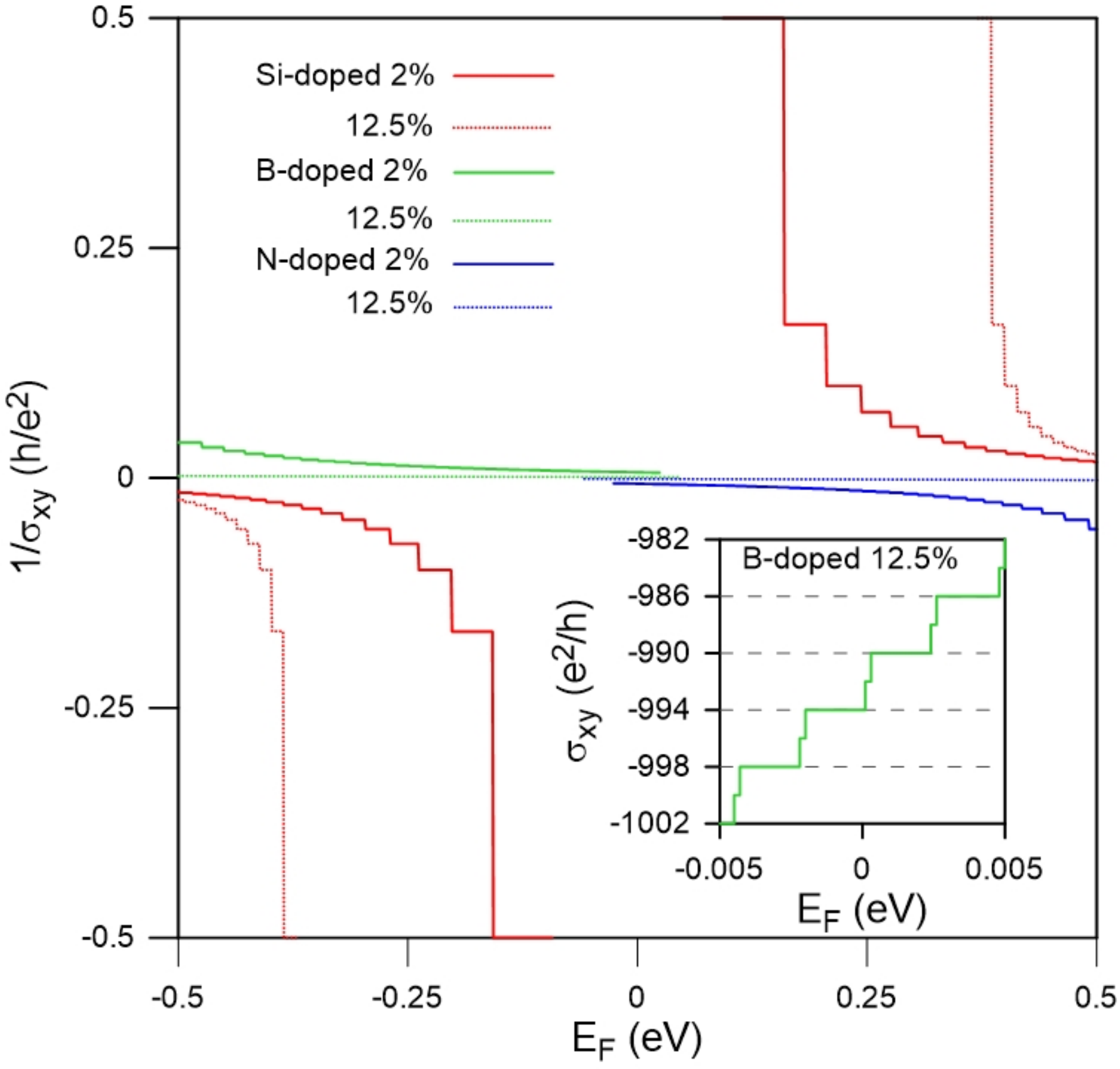}}
\caption{(Color online) The $E_F$-dependent resistivity of Si-, B-, and N-doped graphene for $B_z$ = 20 T. The solid and dashed curves represent the dopant densities of 2\% and 12.5\%, respectively. The inset shows the quantum Hall conductivity of 12.5\% B-doped graphene. }
\label{Figure 5}
\end{figure}

\medskip
\par

The conductivity is quantized following a special sequence of $\sigma_{xy} = \{ 0; 2(2m+1) e^2/h \} $ ($m$ is an integer) for all dopants. It should be noticed that, the $E_F$ on the x-axis in Fig. 5 represents the shift of energy from the original Fermi level. The band gap created by the doping is reflected in the discontinuity of $R_{xy}$ near the zero energy, as shown for Si guest atom by the solid and dashed red curves. This is associated with the emergence of zero conductivity, clearly demonstrated in Figs. S3(a) and S3(d) of the Supplemental Materials. As a result, there exist two half-integer steps of $2e^2/h$ height near zero energy, in addition to the conventional steps of $4e^2/h$. This is in contrast with that for pristine graphene where the two half-integer steps at zero energy are not separated.\cite{5} The width of the plateau at zero energy, which directly reflects the size of the band gap, strongly depends on the dopants and doping concentrations.

\medskip
\par

The quantum Hall conductivities in the vicinity of the Fermi levels in the presence of B and N dopants deserve careful scrutiny since they could be identified in magneto-transport measurements. For sufficiently small doping concentration, e.g., 2\% as illustrated by the solid blue and green curves in Fig. 5 (also in Figs. S3(b) and S3(c)), the quantized sequence of quantum Hall conductivity remains unchanged even at higher or lower energy ranges. The doping effect might not be strong enough to remarkably alter the essential physical properties of the system. However by increasing the density of dopants, the ensuing split of LLs near $E_F(B)$ and $E_F(N)$ gives rise to the significant changes in quantum Hall effect. There, the wider and smaller conductivity plateaus are interspersed in the spectra, referring to the inset plot for 12.5\% B-doped graphene (also in Figs. S3(e) and S3(f)), in which the latter is related to the splitting of degenerate LLs. Accordingly, each step is reduced by half in height, obeying a new quantized sequence of $\sigma_{xy} =  2m e^2/h $.

\medskip
\par

Our theoretical predictions for doped graphene of various dopants should provide useful information for further magnetic transport measurements, as done for monolayer and few-layer graphene.\cite{5,6,8,9,10,11} Specifically, we observed the half-integer QHE as predicted to exist in monolayer graphene by different theory groups,\cite{1,2,3} situated on thin graphite films,\cite{4} and verified by experimental measurements.\cite{5,6} Such a special quantization in graphene is attributed to the quantum anomaly of the ${n=0}$ LL corresponding to the Dirac point.\cite{1} The results obtained in this work by the well-developed theoretical framework are expected to be confirmed experimentally.

\section{Concluding Remarks}
\label{sec4}

In conclusion, the generalized tight-binding model and the Kubo formula are employed to explore the interesting QHE in doped graphene for Si, B, and N guest atoms with various concentrations. The theoretical framework takes into consideration substantial factors of the doping system, including the non-uniform bond lengths, site energies, hopping integrals, and external field. This method could be further developed to investigate many other condense-matter materials. The doped graphene is an emergent 2D binary compound material, which presents diverse electronic and transport properties under magnetic quantization. The doping effect can open a band gap of different sizes, leading to remarkable changes in the main features of LLs and thus the magneto-transport properties. The Fermi-energy-dependent QHE exhibits both the integer and half-integer conductivities. The step structure could be controlled by changing the dopants or doping concentration, in which the latter has more significant influence. The theoretical predictions should provide useful information and could be examined by magneto-transport measurements.

\begin{acknowledgements}
The authors thank the Ministry of Science and Technology of Taiwan (R.O.C.) under Grant No. MOST 105-2112-M-017-002-MY2. T. N. Do would like to thank the Ton Duc Thang University.
\end{acknowledgements}

\newpage

$\textbf{References}$


\begin{references}
%mono graphene and AB graphene 1-7
\bibitem{1} V. P. Gusynin and S. G. Sharapov, Phys. Rev. Lett. {\bf 95}, 146801 (2005).

\bibitem{2} N. M. R. Peresab, F. Guineaac, A. H. Castro Netoa, Annals of Physics 321, 1559 (2006).

\bibitem{3} A. Tsukuda, H. Okunaga, D. Nakahara, K. Uchida, T. Konoike and T. Osada, Journal of Physics: Conference Series 334, 012038 (2011).

\bibitem{4} K. S. Novoselov, A. K. Geim, S. V. Morozov, D. Jiang, Y. Zhang, S. V. Dubonos et al., Science 306, 666 (2004).

\bibitem{5} K. S. Novoselov, A. K. Geim, S. V. Morozov, D. Jiang, M. I. Katsnelson, I. V. Grigorieva et al.,  Nature 438, 197 (2005).

\bibitem{6} Y. Zhang, Y. W. Tan, H. L. Stormer and P. Kim,  Nature 438, 201 (2005).

\bibitem{7} E. McCann and V. I. Falko, Phys. Rev. Lett. 96, 086805 (2006).

%trilayer graphene 8-14
\bibitem{8} T. Taychatanapat, K. Watanabe, T. Taniguchi and P. Jarillo-Herrero, Nature Physics 7, 621 (2011).

\bibitem{9} S. H. Jhang, M. F. Craciun, S. Schmidmeier, S. Tokumitsu, S. Russo, M. Yamamoto et al., Phys. Rev. B 84, 161408(R) (2011).

\bibitem{10} L. Zhang, Y. Zhang, J. Camacho, M. Khodas and I. Zaliznyak, Nature Physics 7, 953 (2011).

\bibitem{11} A. Kumar, W. Escoffier, J. M. Poumirol, C. Faugeras, D. P. Arovas, M. M. Fogler et al., Phys. Rev. Lett. 107, 126806 (2011).

\bibitem{12} M. Koshino and E. McCann, Phys. Rev. B 83, 165443 (2011).

\bibitem{13} S. Yuan, R. Roldán, and M. I. Katsnelson, Phys. Rev. B 84, 125455 (2011).

\bibitem{14} T. N. Do, C. P. Chang, P. H. Shih, J. Y. Wu and M. F. Lin, Phys. Chem. Chem. Phys., 19, 29525 (2017).

%others 15-20
\bibitem{15} L. Li, F. Yang, G. J. Ye, Z. Zhang, Z. Zhu, W. Lou et al., Nature Nanotechnology 11, 593–597 (2016).

\bibitem{16} R. S. Meena, Ss Ks Singh, As Pal, As Kumar, R. Jha, K. V. R. Rao et al., Jour. Appl. Phys. 111, 07E323 (2012).

\bibitem{17} Rs Singha, Ss Roy, As Pariari, Bs Satpati, and Ps Mandal, Phys. Rev. B 99, 035110 (2019).

\bibitem{18} F. Yang, Z. Zhang, N. Z. Wang, G. J. Ye, W. Lou, X. Zhou et al., Nano Lett. 18 (10), 6611–6616 (2018).

\bibitem{19} R. Ma, S. W. Liu, W. Q. Yang and M. X. Deng, Europhysics Letters 119, 3 (2017).

\bibitem{20} J. Yin, S. Slizovskiy, Y. Cao, S. Hu, Y. Yang, I. Lobanova et al., Nat. Phys. (2019). DOI: 10.1038/s41567-019-0427-6.

%sts 21
\bibitem{21} D. L. Miller, K. D. Kubista, G. M. Rutter, M. Ruan, W. A. de Heer, P. N. First et al., Science 324 (5929), 924-927 (2009).

%optical spectroscopy 22
\bibitem{22} Z. Jiang, E. A. Henriksen, L. C. Tung, Y. -J. Wang, M. E. Schwartz, M. Y. Han et al., Phys. Rev. Lett. 98 (19), 197403 (2007).

%Si-doped 23
\bibitem{23} W. Zhou, M. D. Kapetanakis, M. P. Prange, S. T. Pantelides,S. J. Pennycook, J. -C. Idrobo, Phys. Rev. Lett. 109 (20), 206803 (2012).

%B-doped 24
\bibitem{24} L. S. Panchakarla, K. S. Subrahmanyam, S. K. Saha, A. Govindaraj, H. R. Krishnamurthy, U. V. Waghmare et al., Advanced Materials 21 (46), 4726-4730 (2009).

\bibitem{B} Z. H. Sheng, H. L. Gao, W. J. Bao, F. B. Wang, and X. H. Xia, J. Mater. Chem. 22, 390 (2012).

%N-doped   25
\bibitem{25} L. Qu, Y. Liu, J. -B. Baek, L. Dai, ACS Nano 4 (3), 1321-1326 (2010).

\bibitem{N} A. Yanilmaz, A. Tomak, B. Akbali, C. Bacaksiz, E. Ozceri, and O. Ari et al., RSC Adv. 7, 28383 (2017).

%non-uniform hopping 26-27
\bibitem{26} S. J. Zhang, S. S. Lin, X. Q. Li, X. Y. Liu, H. A. Wu, W. L. Xu et al., Nanoscale 8 (1), 226-232 (2015).

\bibitem{27} M. Shahrokhi, C. Leonard, Journal of Alloys and Compounds 693, 1185-1196 (2017).

%magnetic quantization Si-doped   28
\bibitem{28} P. H. Shih, T. N. Do, B. L. Huang, G. Gumbs, D. Huang and M. F. Lin, Carbon 144, 608-614 (2019).

%tbm [29]
\bibitem{29} Chen, S.-C.; Wu, J.-Y.;Lin, C.-Y.;Lin, M.-F, IOP Publishing 2053-2563, ISBN: 978-0-7503-1674-3 (2017).

%Kubo formula [30]
\bibitem{30} P. Dutta, S. K. Maiti, and S. N. Karmakar,  Jour. Appl. Phys. \textbf{112}, 044306 (2012).

%gradient approx [31]
\bibitem{31} Lin, M. F.; Shung, K. W.-K, Phys. Rev. B 50 (23), 17744-17747 (1994).

\bibitem{theo1} Y. C. Zhou, H. L. Zhang, and W. Q. Deng Nanotechnology 24, 225705 (2013).

\bibitem{theo2} Y. Fujimoto, Advances in Condensed Matter Physics 2015, 571490 (2015).

\bibitem{theo3} S. S. Varghese, S. Swaminathan, K. K. Singh, and V. Mittal, Electronics 5, 91 (2016).

\end{references}
\end{document}